\documentclass[twoside,leqno,twocolumn]{article}

\usepackage[letterpaper]{geometry}

\usepackage{ltexpprt}
\usepackage{algorithm} 
\usepackage{algpseudocode} 
\usepackage{multirow}
\usepackage{multicol}
\usepackage{graphicx}

\begin{document}

\title{\Large Echelon: Two-Tier Malware Detection for Raw Executables \newline to Reduce False Alarms}
%\author{Authors \thanks{Affiliation}
\author{Anandharaju Durai Raju\thanks{aduraira@sfu.ca, Simon Fraser University, Canada}
\and Ke Wang\thanks{wangk@sfu.ca, Simon Fraser University, Canada}
}

\date{}

\maketitle

% Copyright Statement
% When submitting your final paper to a SIAM proceedings, it is requested that you include 
% the appropriate copyright in the footer of the paper.  The copyright added should be 
% consistent with the copyright selected on the copyright form submitted with the paper.
% Please note that "20XX" should be changed to the year of the meeting.

% Default Copyright Statement
\fancyfoot[R]{\scriptsize{Copyright \textcopyright\ 20XX by SIAM\\
Unauthorized reproduction of this article is prohibited}}

% Depending on which copyright you agree to when you sign the copyright form, the copyright 
% can be changed to one of the following after commenting out the default copyright statement
% above.

%\fancyfoot[R]{\scriptsize{Copyright \textcopyright\ 20XX\\
%Copyright for this paper is retained by authors}}

%\fancyfoot[R]{\scriptsize{Copyright \textcopyright\ 20XX\\
%Copyright retained by principal author's organization}}

%\pagenumbering{arabic}
%\setcounter{page}{1}%Leave this line commented out.

\maketitle

% why called framework
% pre-trained model as first tier

\begin{abstract}
Existing malware detection approaches suffer from a simplistic trade-off between false positive rate (FPR) and true positive rate (TPR) due to a single tier classification approach, where the two measures adversely affect one another. The practical implication for malware detection is that FPR must be kept at an acceptably low level while TPR remains high. To this end, we propose a two-tiered learning,  called \textit{Echelon}, from raw byte data with no need for hand-crafted features. The first tier locks FPR at a specified target level, whereas the second tier improves TPR while maintaining the locked FPR. The core of Echelon lies at extracting activation information of the hidden layers of first tier model for constructing a stronger second tier model. Echelon is a framework in that it allows any existing CNN based model to be adapted in both tiers. 
We present experimental results of evaluating Echelon by adapting the state-of-the-art malware detection model ``Malconv" in the first and second tiers.
\end{abstract}

\section{Introduction}
Malware detection techniques like static and dynamic analysis deal with static codes and dynamic behaviors of malware, respectively. While dynamic analysis excels more than static methods, they are inefficient when used alone. Nowadays, both methods are used to complement each other's advantages in form of hybrid analysis techniques \cite{damodaran2017comparison}. %\citeauthor{damodaran2015combining},
%\cite{eskandari2013hdm}.
%\cite{choi2012toward}
In this paper, we focus on static, end-to-end deep neural network methods, to learn from raw byte sequence data with no need for hand-crafted features, similar to \cite{raffin}\cite{krvcal2018deep}. Another reason for considering raw byte sequences is that such data can be easily obtained without running a portable executable (PE) sample and with little data pre-processing effort.

%\begin{table}[bht!]
%centering
%\begin{tabular}{|p{2.2cm}|p{1cm}|p{1cm}|p{1cm}|} 
%\cline{3-4}
% \multicolumn{2}{c|}{} & \multicolumn{2}{c|}{Actual Class}  % \\ \cline{3-4}
% \multicolumn{2}{c|}{} & \multicolumn{1}{c|}{P (+)} & %\multicolumn{1}{c|}{N (-)} \\ \hline
% \multirow{2}{*}{Predicted Class} & \multicolumn{1}{c|}{P %+)} & \multicolumn{1}{c|}{TP} & \multicolumn{1}{c|}{FP} \\ %cline{2-4}
%                 & \multicolumn{1}{c|}{N (-)} & %\multicolumn{1}{c|}{FN} & \multicolumn{1}{c|}{TN} \\ \hline
%\end{tabular}
%\caption{Confusion Matrix}
%\label{table:1}
%end{table}

\begin{figure}
\centering
\includegraphics[scale=0.35]{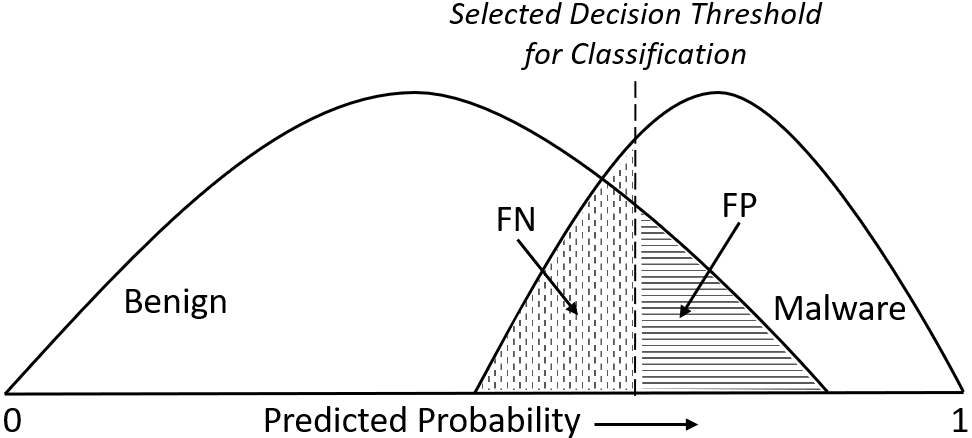} 
\caption{Trade-off between TP and FP}
\label{venn}
\end{figure}

The performance of malware detection software is defined by true positive rate (TPR) and false positive rate (FPR). TPR is the percentage of positive (malware) samples that are detected as positive, and FPR is the percentage of negative (benign) samples that are detected as positive. Using the confusion matrix, $TPR=\frac{TP}{TP+FN}$ and $FPR=\frac{FP}{FP+TN}$. 
% False nagetive rate (FNR) is the percentage of positive samples detected as negative. TPR=1-FNR. 
The cyber-security community is well aware of the importance of having a low FPR because a high FPR leads to many false alarms, which requires excessive manpower to manually investigate on each case, eventually reducing the confidence of using the system. On the other hand, there are different detection models that can be stacked together to catch FNs (i.e., malware files misclassified as benign) \cite{vinayakumar2019robust}\cite{damodaran2017comparison}. 
% To the malware detection community, it is essential to keep false alarms at a very low level.
% For instance, during an anti-malware scan in a PC machine, suppose legitimate critical Windows system driver files are frequently falsely detected as malware and get quarantined or deleted, entire OS might get corrupted and could be more destructive than an actual malware infection.
According to \cite{polyakov2014mitigating},
%\cite{parinov2018elimination}%\cite{101},
it is essential to keep FPR at 1\% or lower 
while TPR remains no less than 90\%. However, achieving such performance is a challenge for traditional single tier classification, such as \cite{krvcal2018deep}\cite{tabish2009malware}\cite{raffin}\cite{le2018deep}, which suffers from the simplistic TPR/FPR trade-off due to the decision threshold shown in Figure \ref{venn} 
- a reduced FPR is always at the expense of a lowered TPR (note $TPR=1-\frac{FN}{FN+TP}=1-FNR$). 

% The proposed Echelon addresses this issue with a two-tiered learning where the first tier locks FPR at a target level and the second tier improves TPR by learning 
% from TN and FN samples of the first tier and their activation information  
% at the hidden layers of first tier. 
% % Such activation information provides 
% % more detailed features for distinguishing TN samples and FN samples. 

To keep FPR at an acceptably low level while maintaining a high TPR, we propose the \textit{locked FPR} requirement and a \textit{two-tiered} solution called Echelon. 
% The name ``Echelon" comes from a military formation where each subsequent level has a different function to overcome the impediments of former levels. 
Echelon learns the model at two tiers. The first tier 
locks a specified \textit{target FPR} but may have a low TPR. 
The second tier is responsible for improving TPR while having no or little increase of FPR. With each tier having its own decision threshold, Echelon does not suffer from the simplistic trade-off of TPR/FPR that results due to a single decision threshold. Echelon is a framework in that it can convert an existing convolutional neural network (CNN) based model into two-tiered models by adapting it in both first and second tiers; the only assumption is that the CNN model 
has a max pooling layer followed by a fully connected network. 

% The novelty lies at the extraction of high activation information of the first tier, from which the second tier can be learnt to improve TPR while preserving the FPR locked by the first tier; this extraction requires only that models are CNN based. 

% More precisely, the first tier locks the target FPR through 
% choosing a decision threshold for the adapted CNN model. 

The challenge lies at learning a strong second tier model that must lock the target FPR achieved by the first tier while substantially increasing the overall TPR. 
%To achieve this goal, we consider convolutional neural networks (CNN) as our model representation. 
We learn the second tier model through a novel \textit{Activation Trend Identification (ATI)} mechanism. ATI focuses on the TN and FN samples 
of the first tier and identifies the PE sections that have high activations at the hidden layers of first tier model for either class but not both. 
Such high activation sections provide sharpened features to distinguish TN samples and FN samples in the second tier, thereby, correcting FNs of the first tier while introducing zero or little new FPs. 

We further consider two enhancements for the above section-based approach. The first enhancement focuses on actual data regions within an identified PE section that yield high activations, instead of whole PE section. This restriction helps reduce the training data size in expensive Tier-2 training. The second enhancement incorporates PE section level semantics, in addition to activation data regions, for utilizing high level information modality offered by different PE sections.

As a proof of concept, we adapted the state-of-the-art CNN malware detection model, Malconv\cite{raffin}, into the first and second tiers of Echelon and evaluated it on 
real malware datasets. The study shows that a significant improvement of up to $\approx5\%$ over Malconv's TPR is achievable while retaining FPR at target 0.1\%. This improvement is obtained by the second tier over difficult samples that Malconv failed to learn and classify.

The rest of the paper is organized as follows. Section \ref{sec:related} reviews related works. In Section \ref{sec:background}, a brief background on PE file format and the basic Malconv architecture are provided. Section \ref{sec:framework} presents our two-tier Echelon framework, followed by the details of activation trend identification in Section \ref{ati} and the training algorithms in Section \ref{sec:training}. Section \ref{sec:evaluation} presents the datasets and experimental results. Finally, Section \ref{sec:conclusion} concludes this study.% and discusses future work.

\section{Related Work}\label{sec:related}
We consider three categories of related works.
% : those that directly work on byte sequences, those that extract and use features from byte sequences, and those that use multiple weak learners to derive a strong overall learner. 

\subsection{Based on Byte Sequences:}
Most works dealing with byte sequence data are based on neural networks. 
Raff et al\cite{raffin} and Coull et al\cite{coull2019activation} are single-tiered approaches, which suffer from the simplistic TPR-FPR trade-off. They use raw bytes as base features and analyze hidden layer activations to showcase that deep neural networks could learn information similar to a malware analyst.
% and they do not explore 
% beyond mapping the hidden activation such as using the difference in activation level to re-classify incorrectly classified malware samples like ours. %and also tend to limit their work only to extract smaller fixed size n-gram  bytes information for further processing. 
The Malconv CNN model by \cite{raffin} treats a raw executable as a single byte sequence on its entirety, ignoring the PE section level semantics on multi-modal contents like image, code, etc., thus, loses the benefits of leveraging semantic information for better detection.
 \cite{coull2019activation} auto-extracted low-level 11-byte feature representations from (100KB) samples that resembles manually-derived features, whereas we use higher level PE section data from samples of size up to 1MB, as our automated feature representation for better generalization. 
 
 \cite{krvcal2018deep} proposed a simpler CNN based model for byte data performing similar to \cite{raffin}, and \cite{vinayakumar2019robust} proposed a hybrid framework with classic and deep learning models along with Malconv variants. Both suffer the same issue as described for \cite{raffin}. %, \cite{llaurado2016convolutional}  
\cite{le2018deep} proposed a method for feature identification without expert domain knowledge. Their method completely ignores PE section level semantics of PE samples as they artificially re-scale the 1-D byte code to a fixed 10k bytes length using OpenCV library. Such lossy data compression is prone to over-fitting.

\cite{llaurado2016convolutional} showed that applying deep learning over disassembled raw binary content has merits in considering the semantic meaning of each byte. They used the PE section's characteristics instead of its binary content and used severely downsampled (32x32) malware image representations as input, which would corrupt PE semantics and would not be scalable for real world samples.

% To the best of our knowledge, existing deep neural network based malware detection has not addressed the ``acceptably low FPR while maximizing TPR" using hidden layer information while processing 1-D/2-D image-like byte sequence representation of PE binaries.

\subsection{Based on Feature Extraction:}
Many works assume a vectorized feature space for input that require expensive domain knowledge based feature extraction. 
\cite{tabish2009malware} and \cite{santos2009n} use 
%and \cite{jain2011byte},
fixed size n-gram byte blocks as feature.
% , for training boosted decision trees, k-nn, Instance Based Learner, Naïve Bayes and AdaBoost1 classifiers.  
While n-gram methods are useful, they are computationally expensive due to the enormous n-grams and might not cover long range dependencies %\cite{bengio2003neural}
%\cite{reddy2006n} 
that are important for our purpose, and they are prone to over-fitting, suffer extreme sparsity, and give diminished returns with more data\cite{raff2018investigation}. 
\cite{anderson2018ember} compared the performance of LightGBM model with Malconv model using features such as raw-byte histogram, byte-entropy histogram, etc. Other feature types include string-2D histogram \cite{saxe2015deep}, GIST-based binary textures \cite{nataraj2011comparative}, Markov n-grams' entropy rate \cite{shafiq2008embedded}, and structural byte entropy graphs \cite{xiao2020malfcs}.
Extracting such features needs specialized feature engineering pipelines.

\subsection{Based on Multi Classifiers:}
% The motivation of our two tier model is different from existing ensemble strategies using multi classifiers for improving classification accuracy \cite{lu2010using}: \textit{Bagging:} equal-weight classifiers are trained on several subsets of data from training set followed by voting.
% % \textbf{Grading:} Combines vote cast by heterogeneous classifiers that are 
% \textit{Boosting:} weighted classifiers iteratively learned from weighted dataset by re-weighting misclassified instances.
% \textit{Voting:} votes cast by heterogeneous classifiers are combined to make the final decision.
% \textit{Stacking:} a sequential classifier learning method %\cite{seewald2003towards}
% using two levels to combine heterogeneous base classifiers, where in level-1 each base classifier predicts the probabilities for all  classes, and in level-2 a meta-classifier is trained from the combined prediction probabilities of the base classifiers and true class. 
% In these methods, each weak learner provides additional information to 
% improve the general classification accuracy. 

Existing ensemble strategies use multi classifiers to improve classification accuracy \cite{lu2010using}, such as \textit{Bagging}, \textit{Boosting}, \textit{Voting}, \textit{Stacking}. In these methods, 
each weak learner provides additional information to derive an improved final classification. In contrast, our two tier Echelon is motivated by the specific malware detection requirement to enforce a target FPR while maximizing TPR, which cannot be achieved by simply improving the general classification accuracy. In our two tier model, each tier is responsible for a different subset of prediction: the Tier-1's prediction of the malware class (both TP and FP) is final; Tier-2's purpose is to reclassify misclassified malwares to increase TPR. 
Such a split of prediction is very different from ensemble algorithms where each weak learner  provides additional information to the final learner but does not make a final prediction.

Using n-grams, \cite{abawajy2014large} proposed a multi-tier ensemble of meta-classifier ensembles and \cite{zhang2007malicious} proposed an ensemble of probabilistic neural networks.  \cite{abawajy2017iterative} and \cite{islam2013multi} used two classifiers in Tier-1 and one classifier in Tier-2, focusing on both FPs and FNs for feature selection in the Tier-2. %; Pajouh et al \cite{pajouh2016two} employ dimension reduction techniques like PCA in second tier to improve classification performance. 
These works address general prediction mismatches, instead of locking a target FPR and improving the TPR, and do not directly work with raw byte sequences, but rely on n-grams like feature extraction. 
% Our method directly works with byte sequences with little data pre-processing or feature extraction. 

% and extracts the activation information from hidden layers of CNN. Both the goal and the methodology are very different.

%Finally, we look at anomaly detection methods that attempt to tackle false alarms, though they deal with network traffic and logs data to detect anomalies and are  different from the byte sequence based malware detection discussed in this paper. Consider network intrusion detection systems (NIDS) in anomaly detection, \cite{tjhai2010preliminary} and \cite{mansour2010filtering} attempted to trim down false positives through self-organising maps (SOM) and Growing Hierarchical SOMs respectively. Unlike our targeted FPR locking, and TPR improvement strategy, they aim to reduce either only the FPs, or both FPs and FNs in a generic way encountering the TPR-FPR trade-off.

The two-level hierarchical associative classifier in \cite{ye2010hierarchical} 
aims for high recall, i.e., TPR, in first level while optimizing precision in the next level. Their work lacks locking FPR at a target level, which is our focus, and their work requires API call based feature extraction.

Finally, there are false alarm reduction related works on anomaly detection \cite{anomaly09} using network intrusion detection systems. Such works are very different from our malware detection scenario dealt here.

\section{Background}\label{sec:background}
We first briefly introduce the file format of PE samples and then, discuss shortly about Malconv \cite{raffin} model that will be adapted as our Tier-1 model in the presentation. 
 
\begin{table}[bht!]
\centering
\begin{tabular}{|p{1.8cm}l|} 
 \hline \small
 \textbf{Section Name} & \small \textbf{Description} \\  
 \hline
 .text & Comprises executable code \\
 \hline
 .data & Holds initialized application data \\
 \hline
 .rdata & Read-only data like strings, constants \\
 \hline
% .pdata & Exception information \\
% \hline
 .edata & Export directory for an application \\
 \hline
 .idata & Import directory {\&} address table \\
 \hline
 .bss & Holds uninitialized application data \\
 \hline
 header & Information about other sections \\
 \hline
 .debug & Deugging related information \\
 \hline
 .reloc & Information on image relocations \\
 \hline
 .rsrc & Module resources like images {\&} icons \\ 
 \hline
\end{tabular}
\caption{Typical sections in a PE sample}
\label{table:1}
\end{table}

\subsection{PE Samples:}
Windows Portable Executable (PE) is a Windows 32/64-bit file format used for DLLs, executable programs, etc. 
Here, we consider a broader level sub-area present in a PE file called ``PE Section'', that provides a logical and physical separation to different parts of the program contained in PE file, and also helps in loading the executable file into memory during execution. Table \ref{table:1} contains the details of some PE sections that are prevalent in most PE samples. Other custom made sections may also be present. For example, there were more than 2,500 distinct PE section names found collectively across the samples in our datasets.  

%The PE Header holds the basic information like the number of PE sections. 
A lookup directory named ``Section-Table"\footnote[1]{docs.microsoft.com/en-us/windows/win32/debug/pe-format
%\#section-table-section-headers
}, located usually after Header data, lists meta information like the PE section's names, its virtual size, and location - in the order of their relative virtual address (RVA). Further details about PE sections can be referred at \cite{bytepointer}. 
There exists no uniformity in terms of a PE section's presence, their sizes across samples, and the order in which they appear. Hence, treating PE byte sequence as a 2-D image like  %\cite{nataraj2011malware}
\cite{wagner2015survey} will lose the PE semantics: a same pixel value will have the same intensity anywhere in an image, but a same byte value in a PE sample will have different meanings in different PE sections. While we focus on Windows PE, our work can be transferrable to other file formats such as Linux (ELF file format) based malware, which are on the rise recently. 

\begin{figure}[bht!]
\centering
\includegraphics[scale=0.35, angle=0]{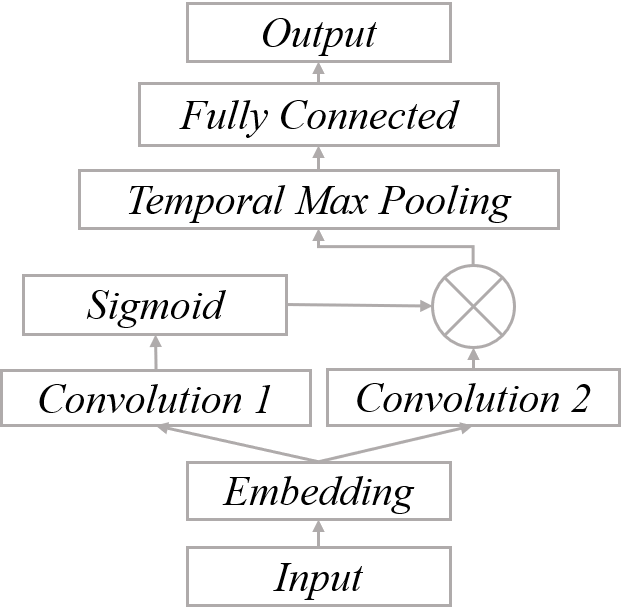} 
\caption{Architecture of Malconv}
\label{malconv_arch}
\end{figure}

\subsection{The Malconv Architecture:}
Our Echelon framework can adapt any CNN-based model with a max pooling followed by a fully connected network. We use Raff et al's Malconv model by \cite{raffin} for our presentation. 
The architecture of the Malconv model is illustrated in Figure \ref{malconv_arch}. It maps input to an 8-dim embedding layer, which is then parallelly fed into two 1-D convolutional layers, one of which is followed by a sigmoid layer. Both layer outputs are multiplied element-wise by coupling them to a gating mechanism, %, which facilitates controlled propagation of only useful information from stacked outputs to further layers and also helps in reducing vanishing gradients. 
followed by a temporal max-pooling layer for obtaining the global maximums, that are fed into a fully connected layer to get final output. More details will be presented in Figure \ref{map}. 

\section{The Echelon Framework}\label{sec:framework}
\subsection{Overview:}
\begin{figure}[bht]
\centering
\includegraphics[scale=0.33]{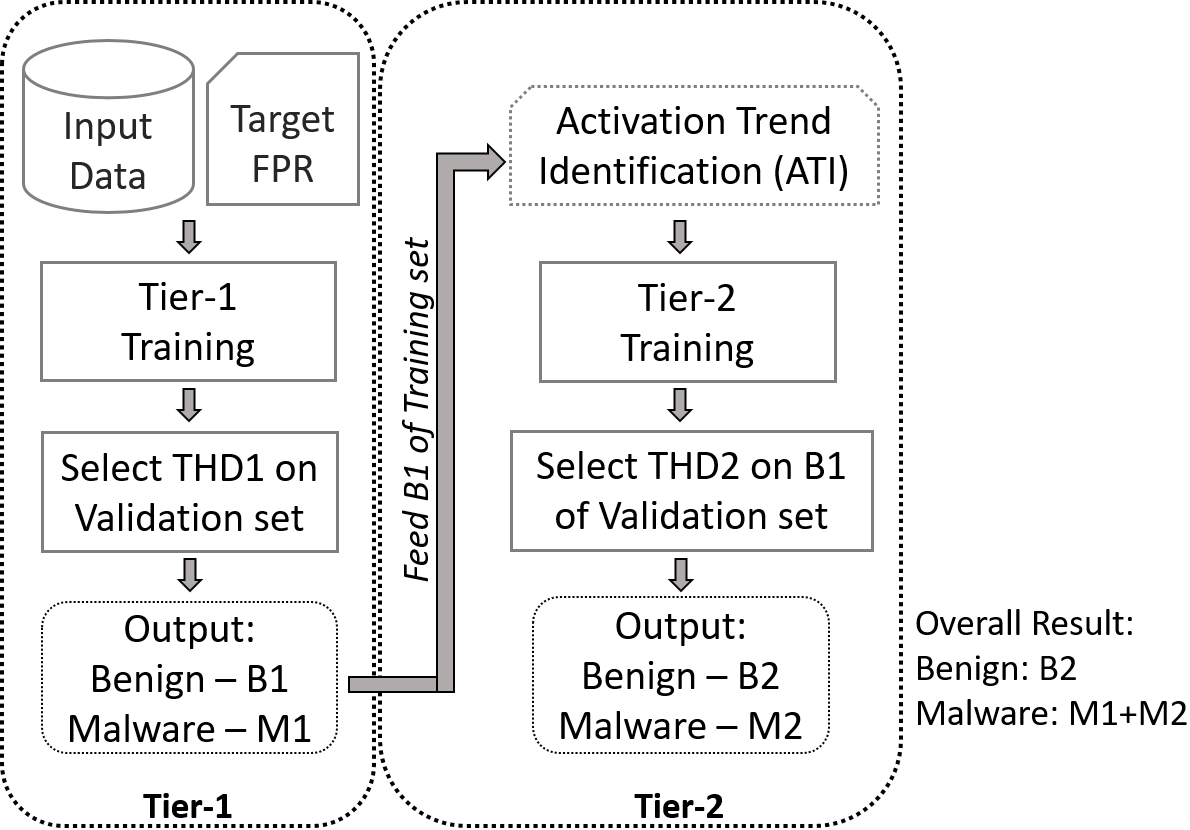} 
\caption{Echelon training flow. %Tier-1 trains on full training set and splits it into $B1_{train}$ and $M1_{train}$, with $THD_1$ chosen to meet the target FPR on validation set. Tier-2 trains on $B1_{train}$ to reclassify the misclassified malware samples in B1, with $THD_2$ selected to have nearly zero FPR on $B1_{val}$.
}
\label{flow}
\end{figure}

The aim of Echelon is to obtain a better TPR while locking FPR at a specified level. It receives a user-specified target FPR as input along with a collection of labeled PE samples. Echelon adapts an existing CNN model and produces two models called Tier-1 model and Tier-2 model, that are trained sequentially to achieve the overall target FPR, and an overall TPR higher than what the Tier-1 model alone could achieve. While the Tier-1 model uses the adapted model, the Tier-2 model has a similar network architecture but 
is trained on a subset of the original training data and selected PE sections or data regions. The 
high level training flow in Echelon is shown in Figure \ref{flow}. Each tier has a specific purpose in the whole malware detection process, described as follows.

\begin{figure*}[bht!]
\includegraphics[scale=0.4]{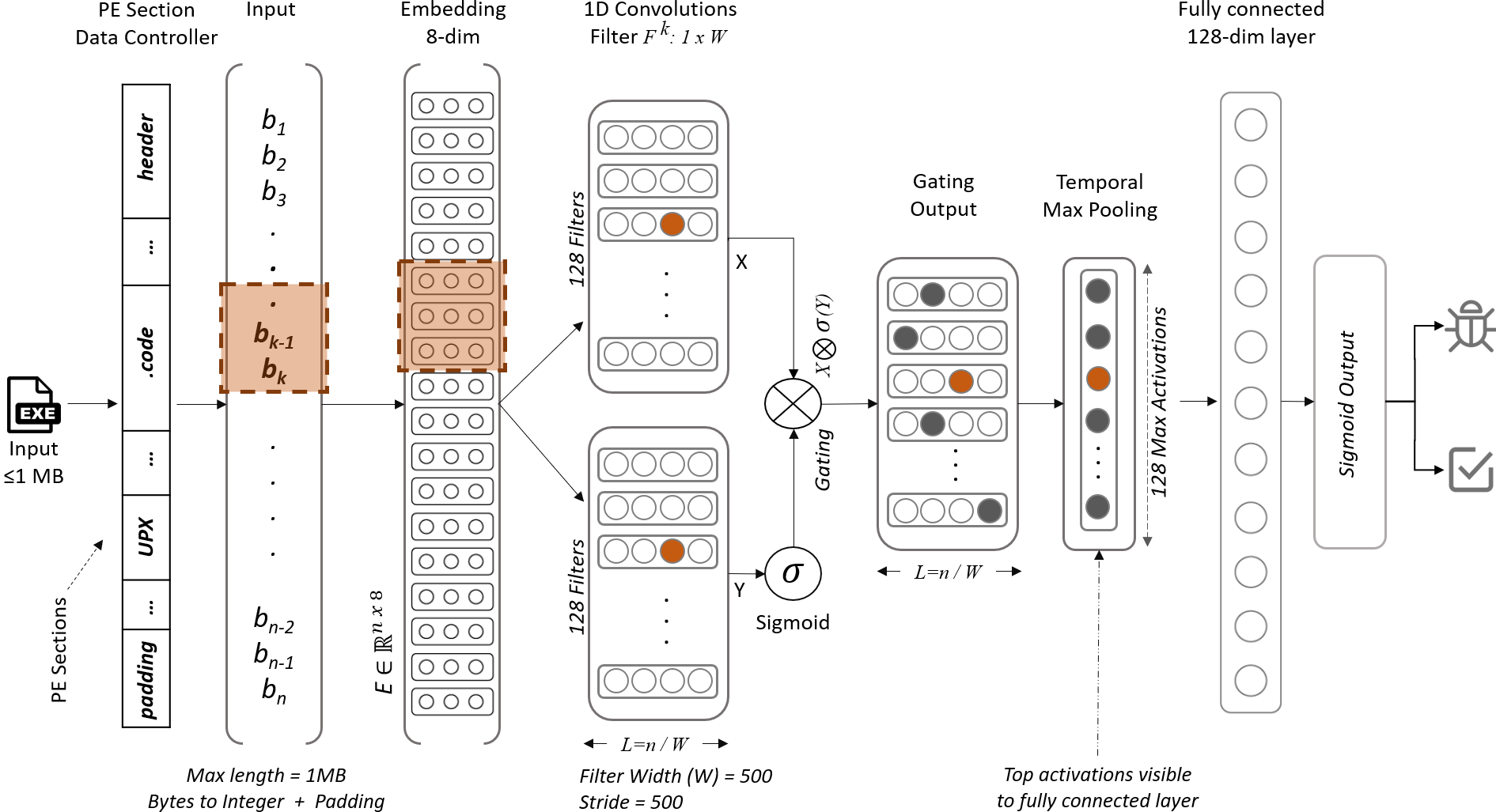} 
\caption{A top activation value generated by max pooling in the Tier-1 model (Malconv). The highlighted places represent the intermediate data responsible for generating the top activation at various stages of the model.}
\label{map}
\end{figure*}

\subsection{Tier-1:}\label{sec:tier-1}
We divide the labeled dataset into training, validation, and testing. Training the Tier-1 model is the same as training the adapted CNN model except that the decision threshold $THD_1$ is selected to comply with the given target FPR on the validation set. The validation set, not the training set, is needed to avoid overfitting. The Tier-1 model classifies the training set into $B1_{train}$ and $M1_{train}$, representing the subset of predicted benign samples and the subset of predicted malware samples, and similarly classifies the validation set into $B1_{val}$ and $M1_{val}$ sets. Note that generically $M1$ consists of true positives ($TP_1$) and false positives ($FP_1$), and $B1$ consists of true negatives ($TN_1$) and false negatives ($FN_1$). The compliance with the target FPR implies that the prediction for the samples in $M1$ is final, but B1 may contain excessive actual malware samples, i.e., FNs, which the Tier-1 model failed to learn and classify due to the target FPR compliance.

% The purpose of Tier-2 model is to correctly classify such samples in B1. 

\subsection{Tier-2:}
The objective of Tier-2 training is to correct the FNs in $B1$, i.e., reclassify them into malware while ensuring that the overall FPR of the two tiers together does not exceed the target FPR. 
%One type of information is the sample level prediction probability of the Tier-1 model, which is less useful. 
Two ideas contribute to our Tier-2 training. First, Tier-2 training focuses on $B1_{train}$, i.e., the training samples classified as benign by Tier-1. Second, Tier-2 training focuses on ``biased" PE sections whose hidden layer activation in the Tier-1 model shows a strong bias for positive vs negative samples in $B1_{train}$, thereby helping in capturing FNs in $B1$ while retaining TNs.
Biased PE sections are identified by  \textbf{Activation Trend Identification (ATI)} presented in the next section. 

To preserve the target FPR, the decision threshold for Tier-2 model, $THD_2$, is selected to be the minimum such that the overall FPR (Equation \ref{ofpr}) of the two tiers on $B1_{val}$ is no more than the target FPR. Tier-2 model splits $B1_{val}$ into $B2_{val}$ and $M2_{val}$,  
%Recall that $B1_{val}$ and $M1_{val}$ denote the subsets of validation samples that are predicted by Tier-1 model as benign and malware, respectively. 
where $B2_{val}$ consists of $TN_2$ and $FN_2$, and $M2_{val}$ consists of $TP_2$ and $FP_2$.
%$M1_{val}$ consists of true positives and false positives, denoted by $TP_1$ and $FP_1$. $B1_{val}$ consists of true positives, false positives, true negatives, false negatives of applying Tier-2 model to $B1_{val}$, denoted by $TP_2,FP_2,TN_2,FN_2$. 
The \textit{overall FPR}  on the validation set
is the percentage of negative samples that are predicted as positive by either Tier-1 model or Tier-2 model: 
\begin{equation}\label{ofpr}
    Overall\_FPR = \frac{FP_1+FP_2}{FP_1+FP_2+TN_2}
\end{equation}
The \textit{overall  TPR} is the percentage of positives that are predicted as positive by either Tier-1 or Tier-2 model:  
\begin{equation}
    Overall\_TPR = \frac{TP_1+TP_2}{TP_1+TP_2+FN_2}
\end{equation}
These metrics are similarly defined for testing set.%, except that the testing set is used instead of the validation set.  

% Let $M1_{val}$ be the set of validation samples predicted as malware by Tier-1 model and let $M2_{val}$ be the set of validation samples in $B1_{val}$ predicted as malware by Tier-2 model. Let $\delta1$ and $\delta2$ be the FPR of $M1_{val}$ and of $M2_{val}$. Then the overall FPR of the two models is $\frac{\delta1|M1|+\delta2|M2|}{|M1|+|M2|}$. 

\subsection{Prediction:}
To classify an unlabeled sample $x$, we apply the two models sequentially, similar to classifying a validation sample. If the Tier-1 model predicts $x$ as malware, i.e., belonging to $M1$, the prediction is final. If the Tier-1 predicts $x$ as benign, i.e., belonging to $B1$, we apply Tier-2 model to predict the final label of $x$, using only the biased PE sections in $x$. 

Next, we discuss how biased PE sections are identified using hidden layer information of the Tier-1 model.

\begin{figure*}[bht!]
\centering
\includegraphics[scale=0.44]{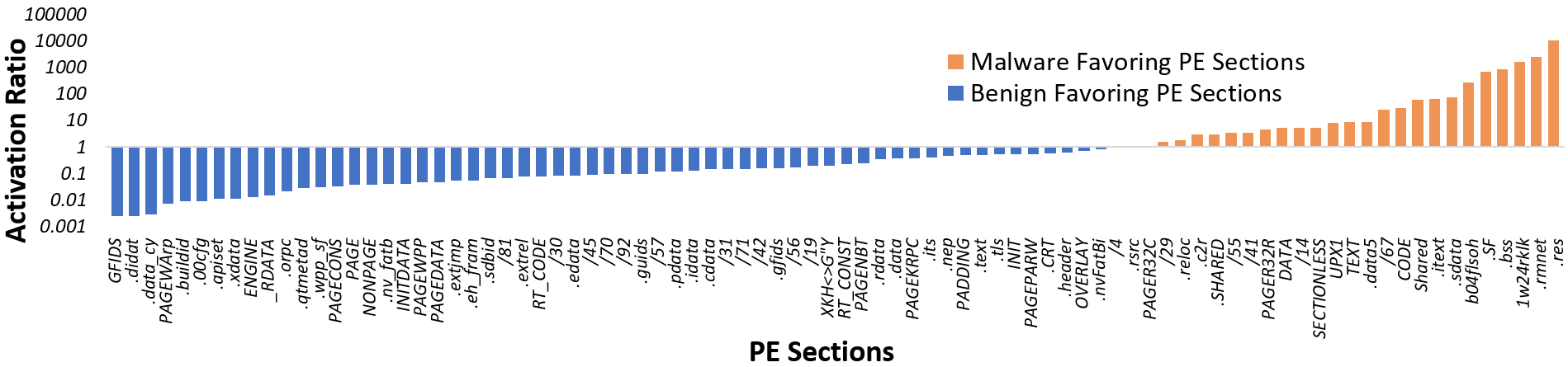} 
\caption{Histogram of $AR_s$ for some PE sections in DS1 dataset.}
\label{mta} 
\end{figure*}

\section{Activation Trend Identification (ATI)} \label{ati}
The functionality of ATI is to identify the PE sections with a strong activation bias in the two classes.
To explain this functionality, we zoom-in the Malconv network in  Figure \ref{map} as the adopted model for our Tier-1 model.  
While there are possibly multiple hidden layers in the Tier-1 model, only the maximum activation that passes through the max pooling is visible to the subsequent fully connected network. Therefore, we shall focus on the \textit{top activations} that survived the max pooling and identify the PE sections that yielded top activations at vastly different rates for benign and malware classes, to train the Tier-2 model. 

Figure \ref{map} illustrates the tracing of a PE section responsible for generating a top activation. From left, we have the byte sequence input, a controller to filter biased PE section data, embedding, and convolution layers with 128 filters. Each filter has the convolutional width $W$ and produces $L=n/W$ activation values on a sample of $n$ bytes. The tracing goes in the opposite direction starting from a top activation value after max pooling. The highlighted places show the corresponding data of the top activation at various stages of the model, which traces back to a portion of $W$ byte input sequence. This relationship allows us to identify the PE section containing these $W$ bytes for each top activation, with the help of a python library called ``pefile" that provides address offsets of PE sections in a PE sample.

Let $S$ be the set of all distinct PE section names. Consider a PE section name $s \in S$ and the benign class denoted by $C^-$. Let $n_{C^-}$ be the total number of $C^-$ samples in $B1_{train}$, $k_s$ be the number of top activations produced by $s$ in the $k^{th}$ sample in $B1_{train}$. We define the \textit{mean benign activation trend} of $s$ as the average number of top activations of $s$ in all the $B1_{train}$ samples:
\begin{equation} \label{mean}
    T_s^- = \sum_{k=1}^{n_{C^-}} k_s \bigg/ n_{C^-}
\end{equation}
Intuitively, $T_s^-$ measures the activation level of the sections named $s$ among benign training samples. Similarly, $T_s^+$ denotes \textit{mean malware activation trend} for sections named $s$  in the malware class $C^+$. We define the \textit{activation ratio} or AR, for each $s$ in $S$ as,
\begin{equation} \label{atgs}
    {AR_s} = T^+_s / T^-_s
\end{equation}

We are interested in the section names $s$ among $S$ with $AR_s$ \textit{far away} from 1, that is, $s$ produces far more top activation in benign samples than in malware samples, or vice versa. This difference makes $s$ an ideal candidate for discriminating between the two classes. 
Figure \ref{mta} shows the histogram for $AR_s$ of some PE section names found in our dataset, sorted by increasing $AR_s$ with blue for $AR_s<1$ and orange for $AR_s>1$. Those sections with a large blue bar or orange bar are biased towards benign or malware, therefore, are useful for training the Tier-2 model. 
Note that large bars of both colors are needed because we need to keep new FPs low while finding many new TPs in Tier-2.  

Let $S_{bias}$ denote a set of section names from $S$ that have a large bar of either color, where ``large" can be defined by some cut-off value either on the bar size or on the number of section names to be included from both sides of the $AR_s$ histogram. In our algorithms below, this cut-off is treated as a hyperparameter for Tier-2 training and is searched using the B1 of validation set.

%The optimal $S_{bias}$ can be searched by treating this $AR_s$ cut-off as a hyperparameter during the Tier-2 training.
% The ``large" bars can be measured in many different ways, such as a threshold on the size of bars, a number of sections from the two ends to be included, etc. The choice of such thresholds and numbers can be searched as a hyperparameter using the validation set. 

\section{Training Algorithms}\label{sec:training}
We now put the above idea in several implementations, where each next implementation improves the previous one by either reducing the training data size or capturing more information for better accuracy.

\subsection{Section-based:}
This is the straightforward implementation, summarized in Algorithm \ref{tpalg}. Tier-1 training is as in Section \ref{sec:tier-1}. The Tier-2 model's architecture is similar to that of Tier-1 model. 
There are two key differences in Tier-2 training: it uses $B1_{train}$ as the training set and keeps only the sections with their names in $S_{bias}$ for Tier-2 training (by removing other PE sections' data from the sample and concatenating remaining $S_{bias}$ section data after padding them
to the nearest multiple of the filter size $W$). 
The same restriction on $S_{bias}$ will apply to the validation set and future prediction. 
The decision threshold $THD_2$ for Tier-2 model is selected to maximize the TPR on $B1_{val}$ subject to the constraint that the overall FPR is no more than the target FPR.
Finally, the Tier-1 model and the best Tier-2 model, together with their $THD_1$ and $THD_2$, are returned. Note that Tier-2 training is more time-consuming than Tier-1 training due to the search for optimal $S_{bias}$ through multiple cut-off values. There is a trade-off between the number of cut-offs considered and Tier-2 training time.

\begin{algorithm}
    \small
	\caption{Two-Tier Training Process} 
	\label{tpalg}
	\begin{algorithmic}[1]
        \Statex \textbf{\textit{Tier-1 Training}}
		\State Train Tier-1 model on training data 
        \State Find $THD_1$ that meets the target FPR on validation set
        \State Let $B1_{train}$ and $B1_{val}$ be the subsets of training set and validation set predicted as benign
        \State Compute $AR_s$ for each PE section $s$ over $B1_{train}$
        \vspace{0.2cm}
        \Statex \textbf{\textit{Tier-2 Training}}
        \ForAll{cut-offs for $S_{bias}$} 
          \State Let $S_{bias}$ = PE section names selected by cut-off 
          \State Train Tier-2 model using $B1_{train}$'s $S_{bias}$ data
          \State Find $THD_2$ with highest TPR on $B1_{val}$ without violating the target FPR by overall FPR
          \State Set $Best$ as Tier-2 model with highest TPR on $B1_{val}$  
     \EndFor
     \State Return Tier-1 and $Best$ Tier-2 models with respective $THD_1$ and $THD_2$
	\end{algorithmic} 
\end{algorithm}

\subsection{Block-based:}
In section-based implementation, the entire data of a $S_{bias}$ PE section is kept to train the Tier-2 model even though a small portion may be responsible for producing the top activation for Tier-1 model. In the block-based implementation, for each top activation from the Tier-1 model, we keep only the portion of data, called a \textit{block}, with a size equal to $W$, which actually generates the top activation. We remove the rest of the bytes in that PE section and concatenate the remaining byte blocks. Then Tier-2 model is trained using such a compressed version of $B1_{train}$ samples. The rest of Algorithm \ref{tpalg} remains unchanged. With our datasets and 
$W=500$, this compression leads to $\approx 94\%$ data reduction compared with $B1_{train}$'s original size. Similar compression is performed for validation and testing/future sets.

% While being large enough to avoid overfitting, the compression also slightly improves model performance than Section-based, as updates to the convolution filters' weight during Tier-2 model training will now be from less noisy bytes. %does not affect the training because only unused bytes are removed.

\subsection{Semantic-Aware:} In Figure \ref{map}, beyond the temporal max pooling layer of block-based implementation, the subsequent fully connected network training is oblivious as to which PE section triggered the top activation. As a result, similar activation values produced by different PE sections have no discriminating power of two classes, even though their PE section names, which represents different semantics, could discriminate the two classes.  To address this issue, we assemble a vector of size same as the activation vector, to provide the corresponding section name for each top activation to the subsequent fully connected network. In Figure \ref{map}, this means that 
the total input size of the fully connected neural network is 256, instead of 128.

\section{Evaluation}\label{sec:evaluation}
%\section{Experiments}\label{experiments}
We provide evidences for our 2-tier approach based on experiments adapting Malconv\cite{raffin} model. 
All experiments were based on the (80:20):20 splits of (Training: Validation): Testing.
%Since, we used pretrained Malconv model as plugin, it required very few epochs to converge, with 2.5 hours per epoch while it took less than 1/5th of the same time for Tier-2 training epoch. 
We set the convolution filter width $W=500$ as used by \cite{raffin} and used a training batch size of 64 samples.  $S_{bias}$ contains an equal number of PE section names from both ends of the activation ratio histogram like in Figure \ref{mta}, where the number is treated as a hyperparameter in Tier-2 training. All the experiments were run on 4x16GB P100 GPUs, with a limit of 200 epochs, an early stopping patience set to 5 epochs, and a learning rate of 0.001 using Adam optimizer.

\begin{table}[bht]
\small
\centering
\begin{tabular}{| p{0.8cm} | p{0.8cm}p{1.cm}p{1.cm} | p{0.7cm}p{0.5cm}p{0.7cm}|} 
 \hline
 &&\textbf{Category}&& Sample&Size& (\footnotesize MB) \\
 \hline
  \textbf{Data}  &  \textbf{Benign} &  \textbf{Malware} &  \textbf{Total} &  \textbf{Min} &  \textbf{Max} &  \textbf{Avg}\\ [0.ex]\hline
 DS1 & 102,275 & 102,272 & 204,547 & 0.0006 & 1 & 0.41 \\\hline
 DS2 & 80,565 & 51,841 & 132,406 & 0.0007 & 1 & 0.33 \\\hline
\end{tabular}
\caption{Dataset Distribution}
\label{dist}
\end{table}

\subsection{Datasets:}
% Due to the known issue of unavailability of benchmark datasets especially for raw PE samples, 
We prepared two datasets of different ratio of the two classes, described as DS1 and DS2 in Table \ref{dist}. 
DS1 was created with $n_{C^+} \approx n_{C^-}$, similar to \cite{raffin}\cite{coull2019activation}, and DS2 with $n_{C^-} >  n_{C^+}$. DS1 comprises the malware corpus provided by VirusTotal\cite{virustotal} falling in the range of Jan to Mar 2019, and DS2 comprises malware corpus from VirusShare\cite{virusshare}, from sets 350 to 356. While it is possible to get these datasets from \cite{virustotal}\cite{virusshare}, their policy\footnote[2]{support.virustotal.com/hc/en-us/articles/115002168385-Privacy-Policy}, however, does not allow us to release their datasets. The benign samples were collected from Windows machines with fresh Windows OS installation of versions 7, 8, and 10. Only the samples with size $\le$ 1MB were selected from the collected corpus to create the datasets. 
%Further experiments on severely imbalanced dataset with 10:3 ratio of $n_{C^-}$:$n_{C^+}$
DS1 and DS2 contained an overall of 2,678 and 2,220 distinct PE section names respectively.
% , among which their $B1_{train}$ set contained 621 and 489 PE sections respectively, which are in turn utilized in the ATI process. 

We report the performance on TPR and FPR using 5-fold CV with the target FPR set at 0.1\% (see a discussion on setting a stringent target FPR in Introduction). An important point to note is that the reported TPR/FPR performance is expected to hold when applying the models to future data where the ratio of the two classes may be different from our testing set. This is because, unlike Accuracy and Precision metrics, TPR measures the probability of detecting malware (positive) samples within the malware class, and this probability is determined by the model trained, not by the relative size of the two classes in testing/future data. The same argument applies to FPR. Hence, the reported TPR/FPR performance will hold on any testing/future sets that have a different $n_{C^-}: n_{C^+}$ class ratio from the one used here. 
Indeed, this is confirmed on testing sets with class ratios 1:1, 5:4, 5:3, 5:2, 5:1, and 10:1.

% and no large files were truncated. % to form the 1MB data. 

%DS1 was created with the number of malware samples equal to the number of benign samples. DS2 was created so that there are much more benign samples than malware samples. 
%Total number of PE sections found in DS1 and DS2 are 2,795 and 2,220 respectively. 

% ($n >> m$), to test our approach in the presence of class imbalance. 

% \subsubsection{\textbf{Combination of Block-based + Section ID implementation:}}
% We experimented combining these methods by making use of both top activation block data and the corresponding section ID vector for each $B1_{train}$ sample as in section-id method. We expected to bring out the advantages from both methods. But at the fully connected layer, as the section-id vector of old top activations, may no longer be related to the new top activations produced over the top activation block data, the model struggled to learn and converge fast during the training phase. 

\begin{table*}[bht]
\footnotesize
\centering
\begin{tabular}{p{4.3cm} | p{1cm} | p{1.1cm} | p{1cm} | p{1cm} | p{1cm} | p{1cm} | p{1.7cm} | p{1.5cm} } 
 
 \multicolumn{1}{p{4.3cm}|}{\small Implementation Type}
 & \footnotesize \textbf{Tier-1 TPR\%} 
 & \footnotesize \textbf{Tier-1 FPR\% }
 & \footnotesize \textbf{Overall TPR\%} 
 & \footnotesize \textbf{Overall FPR\%}
 & \multicolumn{1}{p{1.2cm}|}{\footnotesize New TPs \newline in Tier-2}
 & \multicolumn{1}{p{1.2cm}|}{\footnotesize New FPs \newline in Tier-2}
 & \#. of $S_{bias}$ sections used 
  & \multicolumn{1}{p{1.85cm}}{\footnotesize Tier-1+Tier-2 \newline Training (hrs)} \\ [0.1ex]
 \hline
  \textbf{Section-based 2-tier}                                    & 89.03 & 0.08 & 92.47 & 0.12 &  702.3 &    7.1 &   306.4/621 &   8.7+ 27.1 \\
  \textbf{Block-based 2-tier}                                      & 89.03 & 0.08 & 92.88 & 0.10 &  787.8 &    3.3 &   306.4/621 &   8.7+ 16.5 \\
  \textbf{Semantic-aware 2-tier}                                 & \textbf{89.03} & \textbf{0.08} & \textbf{93.97} & \textbf{0.09} & 1010.4 &    2.1 &   306.4/621 &    8.7+ 15.4 \\ [0.1ex] \hline
  Section-based using all sections                 & 89.03 & 0.08 & 92.56 & 4.66 &  722.7 &  936.6 &   621/621 &    8.7+ 31.4 \\
  Block-based using all sections                   & 89.03 & 0.08 & 92.71 & 3.01 &  754.7 &  599.2 &   621/621 &    8.7+ 18.2 \\
  Semantic-aware using all sections              & 89.03 & 0.08 & 92.76 & 2.87 &  762.9 &  570.6 &   621/621 &    8.7+ 18.1 \\ [0.1ex] \hline
 % Section-based 2-tier using unqualified sections        & 89.03 & 0.08 & 90.35 & 7.12 &    0 &    0 &   x &    9 \\
 % Block-based 2-tier using unqualified sections          & 89.03 & 0.08 & 90.46 & 7.66 &    0 &    0 &   x &    9 \\
 % Section-ID Aware 2-tier using unqualified sections     & 89.03 & 0.08 & 90.76 & 8.04 &    0 &    0 &   x &    9 \\ \hline
  Malconv \cite{raffin}                                   & - 	   & -    & 89.03 & 0.08 &    - &    - &   - &    8.7 \\
  Kr{\v{c}}{\'a}l \textit{et al} \cite{krvcal2018deep}    & - 	   & -    & 86.22 & 0.09 &    - &    - &   - &    13.2  \\
 % Modified Le et al \cite{le2018deep} 				   & - 	   & -    & 90.13 & 0.08 &    - &    - &   - &    - \\
  CNN-Xgboost \cite{ren2017novel}                         & - 	   & - 	  & 93.90 & 2.20 &    - &    - &   - &    9.2 
\end{tabular}

\caption{5-Fold CV results on hold-out testing set over DS1 dataset @ 0.1\% target FPR. Adapted model: Malconv}
\label{consolidated}
% \end{table*}

%\begin{table*}[bht]
%\small
%\centering
%\begin{tabular}{p{0.8cm} | p{1.5cm} | p{1.5cm} | p{1.5cm} | p{1.5cm} | p{1.6cm} | p{1.5cm} | p{1.5cm} | p{1.5cm} } 
% \footnotesize \textbf{Fold}
% & \footnotesize \textbf{Tier-1 TN}
% & \footnotesize \textbf{Tier-1 FP}
% & \footnotesize \textbf{Tier-1 FN} 
% & \footnotesize \textbf{Tier-1 TP}
% & \footnotesize\textbf{Overall TN} 
% & \footnotesize \textbf{Overall FP}
% & \footnotesize\textbf{Overall FN} 
% & \footnotesize \textbf{Overall TP} \\ [0.1ex]
% \hline
%  1 & 20439 & 16 & 2244 & 18210 & 308.4 & 853.5 &  489489 &   6.5 \\
%  2 & 86.55 & 0.09 & 90.76 & 4.87 & 436.9 & 770.1 &  489489 &   6.5 \\
%  3 & 86.55 & 0.09 & 90.78 & 4.85 & 438.1 & 766.2 &  489489 &   6.5 \\
%  4 & 86.55 & 0.09 & 90.76 & 4.87 & 436.9 & 770.1 &  489489 &   6.5 \\
%  5 & 86.55 & 0.09 & 90.78 & 4.85 & 438.1 & 766.2 &  489489 &   6.5 \\ [0.1ex] %\hline
%end{tabular}
%\caption{Confusion matrix of 5-fold Cross Validation for Section ID Aware implementation on Testing set}
%\label{cmatrix}
%\end{table*}
\vspace{0.5cm}
% \begin{table*}[bht]
\footnotesize
\centering
\begin{tabular}{p{4.3cm} | p{1cm} | p{1.1cm} | p{1cm} | p{1cm} | p{1cm} | p{1cm} | p{1.7cm} | p{1.5cm} } 
 
 \multicolumn{1}{p{4.3cm}|}{\small Implementation Type}
 & \footnotesize \textbf{Tier-1 TPR\%} 
 & \footnotesize \textbf{Tier-1 FPR\% }
 & \footnotesize\textbf{Overall TPR\%} 
 & \footnotesize \textbf{Overall FPR\%} 
 & \multicolumn{1}{p{1.2cm}|}{\footnotesize New TPs \newline in Tier-2}
 & \multicolumn{1}{p{1.2cm}|}{\footnotesize New FPs \newline in Tier-2}
 & \#. of $S_{bias}$ sections used 
  & \multicolumn{1}{p{1.85cm}}{\footnotesize Tier-1+Tier-2 \newline Training (hrs)} \\ [0.1ex]
 \hline
  \textbf{Section-based 2-tier}                                    & 86.55 & 0.09 & 89.91 & 0.13 & 351.7 &    6.2 &  198.9/489 &  6.5+ 13.1 \\
  \textbf{Block-based 2-tier}                                      & 86.55 & 0.09 & 90.87 & 0.10 & 445.3 &    2.3 &  198.9/489 &  6.5+ 7.6 \\
  \textbf{Semantic-aware 2-tier}                                 & \textbf{86.55} & \textbf{0.09} & \textbf{91.41} & \textbf{0.09} & 503.9 &    1.3 &  198.9/489 &  6.5+ 7.3 \\ [0.1ex] \hline
  Section-based using all sections                 & 86.55 & 0.09 & 89.51 & 5.39 & 308.4 & 853.5 &  489/489 &   6.5+ 15.3 \\
  Block-based using all sections                   & 86.55 & 0.09 & 90.76 & 4.87 & 436.9 & 770.1 &  489/489 &   6.5+ 9.8 \\
  Semantic-aware using all sections              & 86.55 & 0.09 & 90.78 & 4.85 & 438.1 & 766.2 &  489/489 &   6.5+ 9.5 \\ [0.1ex] \hline
 % Section-based 2-tier using unqualified sections         & 88.47 & 0.09 & 91.76 & 7.66 & 542 & 2311 &  x &   8 \\
 % Block-based 2-tier using unqualified sections           & 88.47 & 0.09 & 91.76 & 7.66 & 189 & 887 &  x &   8 \\
 % Section-ID Aware 2-tier using unqualified sections      & 88.47 & 0.09 & 91.76 & 7.66 & 542 & 2311 &  x &   8 \\ \hline
  Malconv \cite{raffin}                                    &  - 	 & -    & 86.55 & 0.09 & -   &    - &  - &   6.5 \\
  Kr{\v{c}}{\'a}l \textit{et al} \cite{krvcal2018deep}     & - 	 & -    & 81.70 & 0.10 & -   &    - &  - &   9.1  \\
  CNN-Xgboost \cite{ren2017novel}                          & - 	 & - 	& 89.13 & 1.24 & -   &    - &  - &   6.9 
\end{tabular}
\caption{5-Fold CV results on hold-out testing set over DS2 dataset @ 0.1\% target FPR. Adapted model: Malconv}
\label{consolidated2}
\end{table*}
\vspace{0.2cm}
\begin{table*}[bht]
\small

\begin{tabular}{ ccc }
\footnotesize
\begin{tabular}{p{1.9cm} | p{1cm} | p{1cm} | p{1cm} | p{1cm} } 
 \footnotesize \textbf{DS1 Dataset}
 & \footnotesize \textbf{TN}
 & \footnotesize \textbf{FP}
 & \footnotesize \textbf{FN} 
 & \footnotesize \textbf{TP} \\ [0.1ex]
 \hline
  Tier-1 & -%20438.6 
  & 16.4 & -%2243.8 
  & 18210.2 \\
  Tier-2 & 20436.5 & \textbf{2.1} & 1233.4 & \textbf{1010.4} \\
  Overall & 20436.5 & 18.5 & 1233.4 & 19220.6 \\\hline 
\end{tabular}&&

\begin{tabular}{p{1.9cm} | p{1cm} | p{1cm} | p{1cm} | p{1cm} } 
 \footnotesize \textbf{DS2 Dataset}
 & \footnotesize \textbf{TN}
 & \footnotesize \textbf{FP}
 & \footnotesize \textbf{FN} 
 & \footnotesize \textbf{TP} \\ [0.1ex]
 \hline
  Tier-1 & -%16099.1 
  & 13.9 & -%1394.5 
  & 8973.5 \\
  Tier-2 & 16097.8 & \textbf{1.3} & 890.6 & \textbf{503.9} \\
  Overall & 16097.8 & 15.2 & 890.6 & 9477.4 \\ \hline
\end{tabular}
\end{tabular}
\caption{Averaged Confusion matrix of 5-fold CV for Semantic-aware implementation. The number of benign and malware samples in DS1 testing set are 20,455 and 20,454 respectively, and in DS2 are 16,113 and 10,368 respectively. Note that Tier-1 has only FP and TP because only the prediction of positive class is final.}
\label{cmatrix}
\end{table*}

\subsection{Results for DS1 dataset:}
The results for DS1 testing set are reported in Table \ref{consolidated}. 
The first three rows are the three implementations of the 2-tier framework in Section \ref{sec:training}. The next three are the counterparts that use all sections 
in $S$ in Tier-2 training instead of only those in $S_{bias}$. 
We compare our overall performance with independent results of three state-of-the-art works: Malconv \cite{raffin}, as it is chosen as our Tier-1, Kr{\v{c}}{\'a}l \textit{et al} \cite{krvcal2018deep}, a byte sequence model similar to Malconv but claimed to be performing better, and CNN-Xgboost \cite{ren2017novel}, a multi-classifier approach combining CNN for dealing with sequence data, and Xgboost for boosting accuracy.

% Image recognition approaches like \cite{ren2017novel},  scale down image representations using image libraries to employ Xgboost. Hence, we provide a comparison study using CNNs as a more logical scale down mechanism over 1-D image like byte sequence representation and learn Xgboost on resulting data. Experiments show that our approach is able to perform better by achieving similar TPR as CNN-Xgboost but with lower FPR.

The columns for Tier-1 represent TPR and FPR of Tier-1 only, which is also the result of Malconv when forced to meet the target FPR. The columns for overall results represent the overall TPR and FPR of the 2-tier approach, therefore, the difference is the contribution of the Tier-2 model over the single tier model. The next set of columns give the New TPs and FPs produced by Tier-2, the data reduction in terms of biased sections in $B1_{train}$, and the training time of Tier-1 and Tier-2. Several observations can be drawn from Table \ref{consolidated}. 

\textbf{Observation 1}: Compared to 
the counterparts of using all sections, focusing on data of the $S_{bias}$ sections (such as CODE, UPX1, etc.,) helped lock the FPR over the ``unseen" Testing set. This is because the $S_{bias}$ sections exhibited a larger bias between benign and malware classes (i.e., their $AR_{s}$ are far away from 1). 
On the other hand, the PE sections with $AR_s$ close to 1 could cause the FPR instability on unseen samples as they are poor at discriminating the two classes. This explains why the target FPR was not met on the testing set by the implementations using all sections, though it was enforced on the validation set during training. This comparison reveals the effectiveness of $S_{bias}$ sections.

\textbf{Observation 2}: The block-based and
semantic-aware implementations achieve 3.85\% and 4.94\% improvement in TPR over the Tier-1 respectively, 
while retaining the FPR within the target 0.1\%. 
This improvement is explained by the new TPs and new FPs columns. For example, Tier-2 of the block-based implementation has new TPs about 787, at the expense of a small increase of only 3 FPs. Due to only a small increase of FP, the target FPR continues to be satisfied whereas TPR increases from 89.03\% to 92.88\%. 
The semantic-aware implementation captures an additional $\approx$223 TPs at an even smaller cost of 2 FP, thanks to the section name information. The detailed confusion matrix on semantic-aware's results is given in Table \ref{cmatrix}. 

%are able to mine out and reclassify considerable number of false negatives. 

\textbf{Observation 3}: The semantic-aware and block-based reports well-reduced time consumption in Tier-2 training compared to section-based. For all implementations, time spent on Tier-2 depends on the step size of the $AR_s$ cut-off hyperparameter, which is set to 2\% here. This time can be further reduced by increasing the step size, i.e., reducing number of cut-offs considered.

\textbf{Observation 4}: Our methods outperform all the external algorithms. Within external algorithms, Malconv's individual performance is better than the rest. Malconv and Kr{\v{c}}{\'a}l \textit{et al} report 89.03\% and 86.22\% TPR respectively when forced to meet the target FPR. CNN-Xgboost \cite{ren2017novel} results are not extrapolated, as it failed to meet the target FPR. Kr{\v{c}}{\'a}l \textit{et al}'s model performed poorly on our dataset because they use a combination of local max-pooling and global average-pooling, that are easily influenced by noisy bytes in the input than global (temporal) max-pooling. 
% Table \ref{cmatrix} provides the averaged confusion matrix for DS2 dataset and the section-name aware method.
 
% In contrast, block-based and section-id aware implementations meet the target FPR, yet achieve a similar or better TPR. 
%The result for filter sizes such as 8 and 16, could help in closely understanding the generalization of smaller feature representations that Coul et al \cite{coull2019activation} attempted (11-byte features) versus the larger ones used in this study.

\subsection{Results for DS2 dataset:}

%The activation trend obtained over imbalanced dataset DS2, is illustrated in Figure \ref{intuition-ds2}. 

The results on DS2 with target FPR as 0.1\%, are found in Table \ref{consolidated2}. Here, semantic-aware has achieved 4.86\% improvement over the Tier-1's TPR, 
while keeping overall FPR at 0.09\%, which is within the target FPR. The Tier-1 TPR and overall TPR are relatively lower than that on the balanced DS1, as the model must learn to inhibit new FP occurrences wherein the training set's benign population is more. Again, \cite{ren2017novel} failed to meet target FPR and \cite{krvcal2018deep} reported a lower TPR than Malconv's.

\section{Conclusion}\label{sec:conclusion}
High false alarms (i.e., FPR) lead to excessive manpower investigation and 
reduction of confidence in using the malware detection system.  
Traditional single tier learning suffers from the simplistic trade-off between TPR and FPR due to a single decision threshold. The proposed Echelon addresses this issue with two-tiered learning where the first tier locks FPR at a target level and the second tier improves TPR by learning from activation information of TN and FN samples at the hidden layers of the first tier. Echelon aims to be a general framework by allowing to adapt an existing CNN model in both tiers.  
% Such activation information provides 
% more detailed features for distinguishing TN samples and FN samples. 
% the activation trend information identified through the novel ``Activation Trend Identification" mechanism, for determining biased sections that could aid in correct re-classification of misclassified samples, which is done 
% by mapping hidden layer activations of the given CNN model to their respective PE sections and identifying PE sections with discriminating activation levels in the two classes. 
% Moreover, Echelon uses higher level PE section information as automated byte-sequence feature representation for better generalization. 
Experimental evaluations on samples collected from benchmark sources supported the design goals of Echelon. With PE sections as an automated byte-sequence feature representation, it also serves to improve the attribution of PE sections towards final classification outcomes. 
%For future work, we plan to extend our framework to other neural networks such as Recurrent Neural Networks.

\section{Acknowledgements:}
We acknowledge Ibrahim AbuAlhaol, Yang Zhou, and Huang Shengqiang for their valuable review and feedback for our work.

\bibliographystyle{siamplain.bst}
\bibliography{ms.bib}

\end{document}

\title{\Large Echelon: Two-Tier Malware Detection for Raw Executables \newline to Reduce False Alarms
 - \textbf{Supplementary Material}}
%\author{Authors \thanks{Affiliation}
\author{Anandharaju Durai Raju\thanks{aduraira@sfu.ca, Simon Fraser University, Canada}
\and Ke Wang\thanks{wangk@sfu.ca, Simon Fraser University, Canada}
}

\date{}

% Copyright Statement
% When submitting your final paper to a SIAM proceedings, it is requested that you include 
% the appropriate copyright in the footer of the paper.  The copyright added should be 
% consistent with the copyright selected on the copyright form submitted with the paper.
% Please note that "20XX" should be changed to the year of the meeting.

% Default Copyright Statement
\fancyfoot[R]{\scriptsize{Copyright \textcopyright\ 20XX by SIAM\\
Unauthorized reproduction of this article is prohibited}}

% Depending on which copyright you agree to when you sign the copyright form, the copyright 
% can be changed to one of the following after commenting out the default copyright statement
% above.

%\fancyfoot[R]{\scriptsize{Copyright \textcopyright\ 20XX\\
%Copyright for this paper is retained by authors}}

%\fancyfoot[R]{\scriptsize{Copyright \textcopyright\ 20XX\\
%Copyright retained by principal author's organization}}

%\pagenumbering{arabic}
%\setcounter{page}{1}%Leave this line commented out.

\maketitle

\section{TPR/FPR Trend:}
Generally, all the section-based, block-based and semantic-aware implementations, follow the principle that, the convolution filter width $W$ of Tier-2 is same the $W$ for Tier-1. 
Specifically, the data reduction in section-id aware and block-based implementations depend on the block-size that is directly controlled by the filter width $W$. \cite{krvcal2018deep} recommended to set filter width in powers of 2, since PE compilers tend to align the beginnings of PE sections within PE sample to multiples of powers of two such as 4096. 

\begin{figure}[bht]
\centering
\includegraphics[scale=0.45]{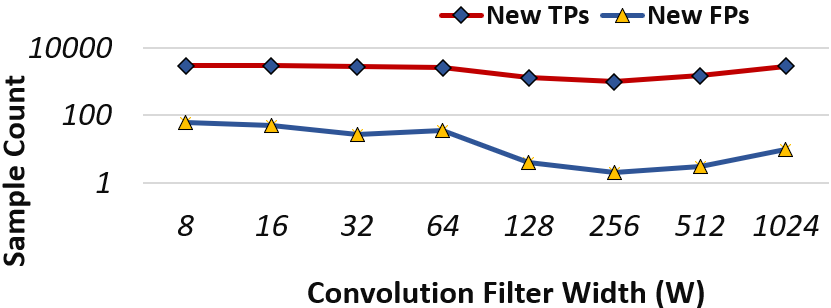} 
\caption{Section-ID Aware implementation: Trend of new TPs/FPs in Tier-2 for different $W$ on DS1 dataset.}
\label{cf}
\end{figure}

\begin{figure}[bht]
\centering
\includegraphics[scale=0.45]{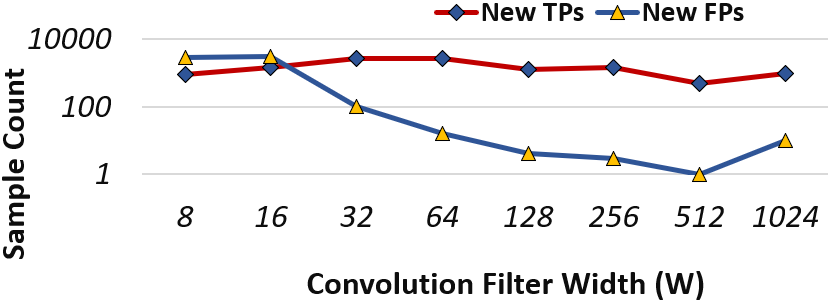} 
\caption{Section-ID Aware implementation: Trend of new TPs/FPs in Tier-2 for different $W$ on DS2 dataset.}
\label{cf2}
\end{figure}

Figures \ref{cf} and \ref{cf2} give the section-id aware implementation results of variation in new TPs/FPs introduced in Tier-2 for different $W$ over DS1 and DS2 repectively. 
The filter widths such as 8, 16 failed to provide enough learnable information to Malconv model, making it to overfit on training data. On the other hand, very large filter widths like 1024 tend to carry noisy information that in turns leads to increased FPR in Tier-2. The optimal block-size found for DS1 is 256 and DS2 is 512.
%  generalization of smaller feature representations that Coul et al \cite{coull2019activation} attempted (11-byte features) versus the larger ones used in this study.

\section{Boosting bound:}\label{b1di}
\begin{figure}[bht]
\centering
\includegraphics[scale=0.32]{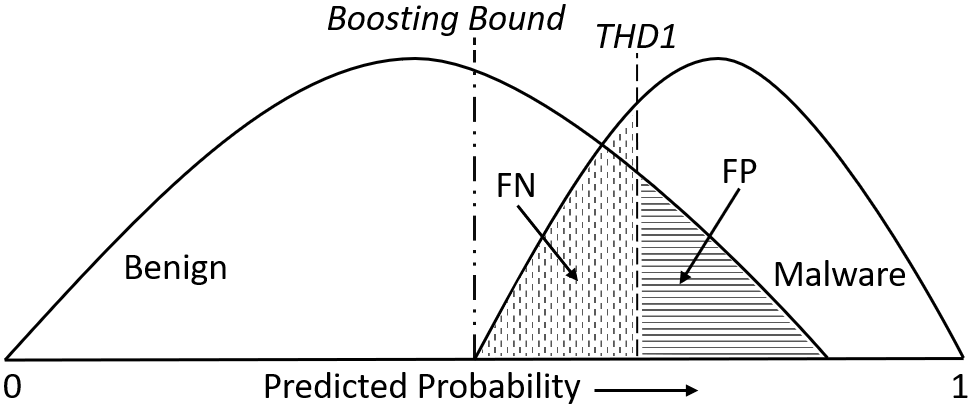} 
\caption{Boosting bound - Samples with output probability less than boosting bound will not be part of B1 set}
\label{bobo}
\end{figure}
Compared to the training set in Tier-1, the distribution of benign and malware samples in the training set for Tier-2, i.e.,  $B1$, is more imbalanced, i.e., $n_{C^+} << n_{C^-}$, because $THD_1$ was set favoring the lowest possible FPR. 
In Figure \ref{bobo}, B1 is the set of samples that are on the left side of $THD_1$, which contains far more benign samples than malware samples, making the Tier-2 training more challenging. To deal with this imbalanced distribution, we can exclude all samples from B1 that are on the left side of the \textit{boosting bound} in Figure \ref{bobo}, which is set to the lowest predicted  malware probability of any malware sample over validation set. 
The reduced B1 contains only the samples with predicted probability of malware class within the range of boosting bound and $THD_1$, which is now more balanced in the two classes, and benefits in speeding up Tier-2 training.

With boosting bound, the following modification is needed for predicting an unlabeled sample. Before applying Tier-2 model, we first check if the predicted probability of malware class is less than boosting bound, if yes, the sample is predicted as benign, otherwise, Tier-2 model is applied to the sample to determine its class. 

\section{Class Population Ratio:}
%Interestingly, our objective of locking FPR and maximizing TPR evaluated on a testing set, will hold on future testing sets (real data) under the constraint of same class distribution but with varying benign to malware class population ratio. 
%Because unlike Precision and Accuracy, metrics like TPR (Recall) and FPR are not affected by the change in relative size of benign and malware class samples - indicating FPR and TPR  performance achieved on a more balanced data will hold on a less balanced real data, provided that both data are sampled from the same distribution. The key point here is that a given classifier is not affected by the choice of testing set.

We make use of TPR and FPR as primary evaluation metrics in our study. It is interesting to examine whether our objective of locking FPR and maximizing TPR evaluated on a testing set, will hold on future testing sets (real data) under the constraint of same class distribution but with varying benign to malware class population ratio. 

Here, we denote class population as the number of samples drawn from a particular class, and class distribution as sample similarity. 

Essentially for a given classifier, TPR and FPR set during training holds even when a different future testing set with different class ratio from that of the actual testing set, is used. Because FPR and TPR are determined by the classifier, not by the testing set. The key point here is that a given classifier is not affected by the choice of testing set.  

By definition of TPR (Recall) and FPR, the portion of samples involved are relative to one class under consideration and are independent of the relative population size of the two classes. In other words, FPR and TPR do not depend on the change in the relative size of malware and benign samples in the testing data and the real data. With this observation,  we can say that the FPR and TPR  performance achieved on a more balanced data will hold on a less balanced real data, provided that both data are sampled from the same distribution. A set of tests done with different testing sets with different class ratio $(n_{C^-}: n_{C^+})$, such as 5:1, 10:1, etc., confirmed the same by resulting in similar TPR-FPR performance.

But other metrics such as Precision: TP/(TP+FP) and Accuracy (TP+TN)/(TP+TN+FP+FN), depends on the relative sizes of malware and benign samples because the underlying TP and FP depend on the two different class sizes. For instance, in case of Precision, by increasing the size of malware files, we have a larger TP and unchanged FP, therefore, a larger TP/(TP+FP). Therefore, though recall does not depend on the relative class size, precision and accuracy does.

\bibliographystyle{siam.bst}
\bibliography{supplementary.bib}